# The CSO Classifier: Ontology-Driven Detection of Research Topics in Scholarly Articles

Angelo A. Salatino, Francesco Osborne, Thiviyan Thanapalasingam, Enrico Motta

Knowledge Media Institute, The Open University, MK7 6AA, Milton Keynes, UK

`{firstname.lastname}@open.ac.uk`

**Abstract.** Classifying research papers according to their research topics is an important task to improve their retrievability, assist the creation of smart analytics, and support a variety of approaches for analysing and making sense of the research environment. In this paper, we present the CSO Classifier, a new unsupervised approach for automatically classifying research papers according to the Computer Science Ontology (CSO), a comprehensive ontology of research areas in the field of Computer Science. The CSO Classifier takes as input the metadata associated with a research paper (title, abstract, keywords) and returns a selection of research concepts drawn from the ontology. The approach was evaluated on a gold standard of manually annotated articles yielding a significant improvement over alternative methods.

## 1 Introduction

Classifying scholarly papers according to the relevant research topics is an important task that enables a multitude of functionalities, such as: (i) categorising proceedings in digital libraries, (ii) enhancing semantically the metadata of scientific publications, (iii) generating recommendations, (iv) producing smart analytics, (v) detecting research trends, and others [1, 2]. Typically, this is done by either classifying the papers in pre-existent categories from domain vocabularies, such as MeSH[1], PhySH[2], and the STW Thesaurus for Economics [3], or by means of topic detection methods, such as probabilistic topic models [3, 4]. The first solution has the significant advantage of relying on a set of formally-defined research topics associated with human readable labels, but requires a good vocabulary of research topics in the domain. Conversely, the latter approaches tend to produce noisier and less interpretable results [5].

We recently released the Computer Science Ontology (CSO) [6], a large-scale, granular, and automatically generated ontology of research areas which includes more than 14K research topics and 162K semantic relationships. CSO has been adopted by Springer Nature editors to classify proceedings in the field of Computer Science, such as the well-known LNCS series [2]. We published this resource to make available to all the relevant communities an open knowledge-base for supporting the development of further applications. However, many users interested in adopting CSO for characterizing their data have limited understanding of semantic technologies and how to use an ontology for annotating documents. Hence, the natural next step was to develop a classifier that allows all the relevant stakeholders to annotate research papers according to CSO.

---

[1] Medical Subject Headings: https://www.nlm.nih.gov/mesh/
[2] PhySH - Physics Subject Headings: https://physh.aps.org
[3] STW Thesaurus for Economics: http://zbw.eu/stw

In this paper, we present the CSO Classifier, a new approach for automatically classifying research papers according to the Computer Science Ontology (CSO). Since the Computer Science Ontology is not yet routinely used by researchers, it is not possible to adopt supervised machine learning algorithms that would require a good number of examples for all the relevant categories. For this reason, we focused instead on an unsupervised solution that does not require such a gold standard. Similarly to Song and Roth [7] and other relevant literature [8], we consider this approach unsupervised because it does not require labelled examples, even if it uses word embeddings produced by processing a large collection of text.

The CSO Classifier takes as input the metadata associated with a scholarly article (usually title, abstract, and keywords) and returns a selection of research topics drawn from CSO. It operates in three steps. First, it finds all topics in the ontology that are explicitly mentioned in the paper. Then it identifies further semantically related topics by utilizing part-of-speech tagging and world embeddings. Finally, it enriches this set by including the super-areas of these topics according to CSO.

The CSO Classifier was evaluated on a gold standard of manually annotated research papers and demonstrated a significant improvement over alternative approaches, such as the classifier previously used by Springer Nature editors to support the annotation of Computer Science proceedings [9].

In summary, our main contributions are:
1. A new unsupervised approach for classifying papers according to the topics in a domain ontology;
2. An application based on this approach which automatically annotates papers with the 14K research topics in CSO;
3. A novel gold standard including 70 papers in the field of "Semantic Web", "Natural Language Processing", and "Data Mining" annotated by 21 domain experts.

The data produced in the evaluation, the Python implementation of the approaches, and the word embeddings are publicly available at http://w3id.org/cso/cso-classifier.

The rest of the paper is organised as follows. In Section 2, we review the literature regarding the topic detection in research papers, pointing out the existing gap. In Section 3, we discuss the Computer Science Ontology. In Section 4 we describe the new approach adopted by the CSO Classifier. In Section 5 we explain how we generated the gold standard and in Section 6 we evaluate the CSO Classifier against several alternative methods. Finally, in Section 7 we summarise the main conclusions and outline future directions of research.

## 2  Literature Review

The task of characterising research papers according to their topics has traditionally been addressed either by using classifiers for assigning to the articles a set of pre-existent categories, or by topic detection methods [3, 4], which generate topics from the text in a bottom-up style.

The first approach has the advantage to produce clean and formally-defined research topics, and thus is usually preferred when a good characterization of the research topics within a domain is available. For instance, Decker [10] introduced an unsupervised approach that generates paper-topic relationships by exploiting keywords and words extracted from the abstracts in order to analyse the trends of topics on different timescales. Mai et al. [11] developed an approach to subject classification using deep learning techniques and they applied it on a set of paper annotated with the STW Thesaurus for Economics (~5K classes) and MeSH (~27K classes). Similarly, Chernyak [12] presented a supervised approach for annotating papers in Computer Science with topics from ACM.

The second class of approaches are based on topics detection methods. One of the first studies to provide a systematic approach to identifying topics was the Topic Detection and Tracking (TDT) program developed by DARPA [13]. In the literature we can find several approaches that apply clustering techniques to identify topics within a collection of scientific documents [3, 14]. Some approaches rely on just one type of information, e.g., citations [15] or titles [11], while other approaches combine multiple types, e.g., abstract and keywords [2, 10], textual content and citation networks [16]. Several other methods exploit Latent Dirichlet Analysis (LDA) [17], which is a three-level hierarchical Bayesian model that retrieves latent patterns in texts, to model their topics [4]. For instance, Griffiths et al. [4] designed a generative model for document collections, the author-topic model, that simultaneously modeled the content of documents and the interests of authors. A main issue of the approaches that rely on LDA is that they represent topics as a distribution of words and it is often tricky to map them to topics in a classification, although some approaches have been proposed to do so [18].

Another set of methods rely just on keywords. For instance, Duvvuru et al. [19] built a network of co-occurring keywords and subsequently perform statistical analysis by calculating degree, strength, clustering coefficient, and end-point degree to identify clusters and associate them to research topics. Some recent approaches use word embeddings, aiming to quantify semantic similarities between words based on their distributional properties in samples of text. For example, Zhang et al. [20] applied K-means on a set of word represented as embeddings. However, all these approaches to topic detection need to generate the topics from scratch rather than exploiting a domain vocabulary or ontology, resulting in noisier and less interpretable results [5].

In sum, we still lack practical unsupervised approaches for classifying papers according to a granular set of topics. Indeed, most available repositories of scholarly articles, such as Scopus[4], Dimensions[5], and Semantic Scholar[6] adopt keywords or use rather coarse-grained representations of research topics. The CSO Classifier was designed to precisely address this gap and enable high quality automatic classification of research papers in the domain of Computer Science.

---

[4] Scopus - https://www.scopus.com
[5] Dimensions - https://www.dimensions.ai
[6] Semantic Scholar - https://www.semanticscholar.org

## 3 The Computer Science Ontology

The Computer Science Ontology is a large-scale ontology of research areas that was automatically generated using the Klink-2 algorithm [21] on a dataset of 16 million publications, mainly in the field of Computer Science [22]. Differently from other solutions available in the state of the art, CSO includes a much larger number of research topics, enabling a granular characterisation of the content of research papers, and it can be easily updated by running Klink-2 on recent corpora of publications.

The current version of CSO[7] includes 14K semantic topics and 162K relationships. The main root is Computer Science; however, the ontology includes also a few secondary roots, such as Linguistics, Geometry, Semantics, and others.

The CSO data model[8] is an extension of SKOS[9]. It includes four main semantic relations:

- *superTopicOf*, which indicates that a topic is a super-area of another one (e.g., Semantic Web is a super-area of Linked Data).
- *relatedEquivalent*, which indicates that two topics can be treated as equivalent for the purpose of exploring research data (e.g., Ontology Matching and Ontology Mapping).
- *contributesTo*, which indicates that the research output of one topic contributes to another.
- *owl:sameAs*, this relation indicates that a research concepts is identical to an external resource. We used DBpedia Spotlight to connect research concepts to DBpedia.

The Computer Science Ontology is available through the CSO Portal[10], a web application that enables users to download, explore, and provide granular feedback on CSO at different levels. Users can use the portal to rate topics and relationships, suggest missing relationships, and visualise sections of the ontology.

CSO currently supports a range of applications including Smart Topic Miner [2], a tool designed to assist the Springer Nature editorial team in classifying proceedings, Smart Book Recommender [23], an ontology-based recommender system for selecting books to market at academic venues, and several others [6]. It has been used in several research efforts and proved to effectively support a wide range of tasks such as forecasting new research topics, exploration of scholarly data, automatic annotation of research papers, detection of research communities, and ontology forecasting. More information about CSO and how it was developed can be found in [6].

## 4 CSO Classifier

The CSO Classifier is a novel application that takes as input the text from abstract, title, and keywords of a research paper and outputs a list of relevant concepts from CSO. It consists of two main components: (i) the syntactic module and (ii) the semantic module. Figure 1 depicts its architecture.

---

[7] CSO is available for download at https://w3id.org/cso/downloads
[8] CSO Data Model - https://cso.kmi.open.ac.uk/schema/cso
[9] SKOS Simple Knowledge Organization System - http://www.w3.org/2004/02/skos
[10] Computer Science Ontology Portal - https://cso.kmi.open.ac.uk

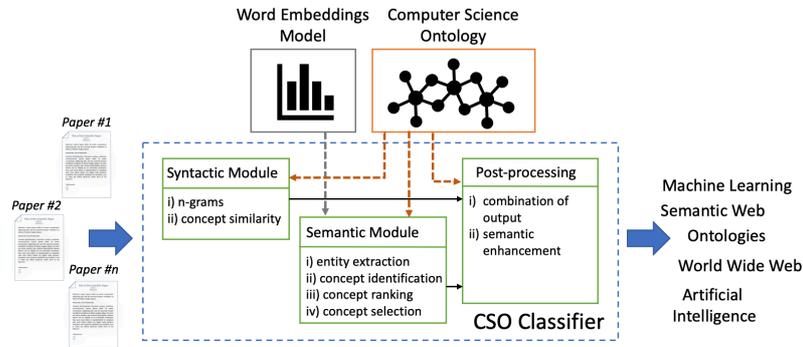

Figure 1. Workflow of the CSO Classifier.

The *syntactic module* parses the input documents and identifies CSO concepts that are explicitly referred in the document. The *semantic module* uses part-of-speech tagging to identify promising terms and then exploits word embeddings to infer semantically related topics. Finally, the CSO Classifier combines the results of these two modules and enhances them by including relevant super-areas. To assist the description of our approach, we will use the sample paper showed in Table 1 [24] as a running example.

Table 1. Sample paper that will be analysed by the CSO Classifier [24].

| **De-anonymizing Social Networks** |
| --- |
| **Authors:** A.Narayanan and V. Shmatikov |
| **Abstract:** Operators of online social networks are increasingly sharing potentially sensitive information about users and their relationships with advertisers, application developers, and data-mining researchers. Privacy is typically protected by anonymization, i.e., removing names, addresses, etc. We present a framework for analyzing privacy and anonymity in social networks and develop a new re-identification algorithm targeting anonymized social-network graphs. To demonstrate its effectiveness on real-world networks, we show that a third of the users who can be verified to have accounts on both Twitter, a popular microblogging service, and Flickr, an online photo-sharing site, can be re-identified in the anonymous Twitter graph with only a 12% error rate. Our de-anonymization algorithm is based purely on the network topology, does not require creation of a large number of dummy "sybil" nodes, is robust to noise and all existing defenses, and works even when the overlap between the target network and the adversary's auxiliary information is small. |
| **Keywords:** social networks, anonymity, privacy |

## 4.1 Syntactic Module

The syntactic module maps n-grams in the text to concepts within CSO. First, the algorithm removes English stop words and collects unigrams, bigrams, and trigrams. Then, for each n-gram, it computes the Levenshtein similarity with the labels of the topics in CSO. Research topics having similarity equal or higher than a threshold (i.e., the constant *msm*) with an n-gram, are selected for the final set of topics, i.e., the *returned topics*. In the prototype *msm* was empirically set to 0.94. This value allows us to recognize many variations of CSO topics and to handle hyphens between words, i.e., "knowledge based systems" and "knowledge-based systems", and plurals, i.e., "database" and "databases".

In Table 2 we report the list of topics returned by the syntactic module for the running example. In contrast with the keyword field, which contains only three terms ("social networks", "anonymity", and "privacy"), the classifier is able to identify a wide range of pertinent topics, such as "microblogging", "data mining", "twitter", and "network topology".

Table 2. Topics returned from the syntactic module when processing the paper in Table 1.

| microblogging, real-world networks, data privacy, sensitive informations, social networks, anonymization, anonymity, online social networks, privacy, twitter, data mining, network topology, graph theory |
|---|

### 4.2 Semantic Module

The semantic module was designed to find topics that are semantically related to the paper but may not be explicitly referred to in it. It utilizes word embeddings produced by word2vec to compute the semantic similarity between the terms in the document and the CSO concepts.

The semantic module follows four steps: (i) entity extraction, (ii) CSO concept identification, (iii) concept ranking, and (iv) concept selection.

In the following sections, we will describe how we trained the word embedding model and illustrate the algorithm.

#### *4.2.1 Word Embedding generation*

We applied the word2vec approach [25, 26] to a collection of text from the Microsoft Academic Graph (MAG)[11] for generating word embeddings. MAG is a scientific knowledge base and a heterogeneous graph containing scientific publication records, citation relationships, authors, institutions, journals, conferences, and fields of study. It is the largest dataset of scholarly data publicly available, and, as of December 2018, it contains more than 210 million publications.

We first downloaded titles, and abstracts of 4,654,062 English papers in the field of Computer Science. Then we pre-processed the data by replacing spaces with underscores in all n-grams matching the CSO topic labels (e.g., "digital libraries" became "digital_libraries") and for frequent bigrams and trigrams (e.g., "highest_accuracies", "highly_cited_journals"). These frequent n-grams were identified by analysing combinations of words that co-occur together, as suggested in [26][12]. Indeed, while it is possible to obtain the vector of a n-gram by averaging the embedding vectors of all its words, the resulting representation usually is not as good as the one obtained by considering the n-gram as a single word during the training phase. Finally, we trained the word2vec model, after testing several combinations of parameters[13].

#### *4.2.2 Entity extraction*

We assume that research concepts can be represented either by nouns or adjectives followed by nouns. Considering only these n-grams reduces the number of text chunks

---

[11] Microsoft Academic Graph - https://www.microsoft.com/en-us/research/project/microsoft-academic-graph/
[12] In particular, for the collocation analysis, we used *min-count* = 5 and *threshold* = 10.
[13] The final parameters of the word2vec model are: *method* = skipgram, *embedding-size* = 128, *window-size* = 10, *min-count-cutoff* = 10, *max-iterations* = 5

to be analysed, speeds up computation and avoids combinations that usually result in false positives. Therefore, the classifier tags each word according to its part of speech (e.g., nouns, verbs, adjectives, adverbs) and then applies a grammar-based chunk parser to identify chunks of words, expressed by the following grammar:

$$<JJ.*>*<NN.*>+ \qquad (1)$$

where JJ represents adjectives and NN represents nouns.

*4.2.3 CSO concept identification*

At this stage, the classifier processes the extracted chunks and uses the word2vec model to identify semantically related topics. First, it decomposes the returned chunks in unigrams, bigrams and trigrams. Then, for each gram, it retrieves from the word2vec model its top 10 similar words (having cosine similarity higher than 0.7). The CSO topics matching these words are added to the result set. Figure 2 illustrates this process more in detail.

When processing bigrams or trigrams, the classifier joins their tokens using an underscore, e.g., "web_application", in order to refer to the corresponding word in the word2vec model. If a n-gram is not available within the vocabulary of the model, the classifier utilizes the average of the embedding vectors of all its tokens.

A specific CSO concept can be identified multiple times due to two main reasons: (i) multiple n-grams can be semantically related to the same CSO concept, and (ii) the same n-gram can appear multiple times within the title, abstract and keywords. For example, the concept "social_media" can be inferred by several semantically related n-grams, such as: "social_networking_sites", "microblogging", "twitter", "blogs", "online_communities", "user-generated_content", and others.

*4.2.4 Concept ranking*

The previous step may produce a large number of topics (typically more then 70), some of which only marginally related to the research paper in question. For instance, when processing the paper in Table 1, some n-grams triggered concepts like "malicious_behaviour" and "gateway_nodes", that may be considered unrelated. For this reason, the semantic module weighs the identified CSO concepts according to their overall relevancy to the paper. The relevance score of a topic is computed as the product between the number of times it was identified (frequency) and the number of unique n-grams that led to it (diversity). For instance, if a concept has been identified five times, from two different n-grams, its final score will equal 10. In addition, if a topic is directly mentioned in the paper, its score is set to the maximum score found. Finally, the classifier ranks the topics according to their relevance score.

*4.2.5 Concept Selection*

The relevance score of the candidate topics typically follow a long-tailed distribution. In order to automatically select only the relevant topics, the classifier adopts the elbow method [27]. This technique was originally designed to find the appropriate number of clusters in a dataset. Specifically, it observes the cost function for varying numbers of clusters. The best number of clusters is then located at the elbow of the resulting curve. This point provides a good trade-off between the number of clusters and the percentage improvement of the cost function.

Figure 3 shows an example of how the elbow method automatically identifies the best cut in the curve of relevance scores, selecting the first 18 topics. In Table 3 we report the list of topics obtained using the semantic module on the running example. In bold are the topics that were detected by the semantic module but not by the syntactic module.

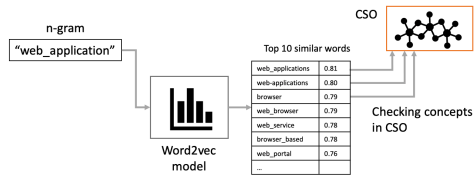

Figure 2. Identification of CSO concepts semantically related to n-grams.

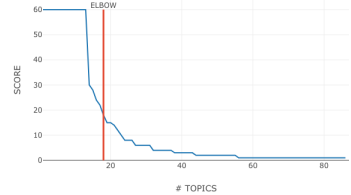

Figure 3. Distribution of the CSO topic scores associated to a paper (blue line), and its elbow (red line).

Table 3. Topics returned from the semantic module when processing the paper in Table 1. In bold the topics missing from the syntactic module in Table 2.

| social networks, anonymity, topology, twitter, anonymization, sensitive informations, data privacy, online social networks, data mining, privacy, **social media**, social networking sites, graph theory, **network architecture**, **micro-blog**, **online communities**, **social graphs** |
|---|

### 4.3 Combined output and enhancement

The CSO Classifier combines the topics returned by the two modules. In this phase, it first discards the topics returned by the semantic module that appear among the first *n* most occurring words in the vocabulary of the embeddings (*n=3,000* in the prototype). This is done because these very generic terms (e.g., 'language', 'learning', 'component') tend to have a good similarity value with a large number of n-grams, typically resulting in too many false positives. It then enriches the combined set of topics by inferring all their direct super topics, exploiting the *superTopicOf* relationship within CSO [6]. For instance, when the classifier extracts the topic "machine learning", it will infer also "artificial intelligence". By default, the CSO Classifier includes only the direct super topics, but it is also possible to infer the list of all their super topics up to the root, i.e., Computer Science.

In Table 4 we report the list of topics inferred from the topics returned by the syntactic and semantic module. As we can see there are several other topics that are pertinent to the paper in Table 1, such as: "security of data", "authentication", "world wide web", and others.

Table 4. Topics obtained from the enhancement process when processing the running paper

| authentication, theoretical computer science, world wide web, privacy preserving, access control, network protocols, complex networks, online systems, network security, security of data, computer science |
|---|

We take the union of the result sets of the two modules, since this solution maximizes the f-measure according to the evaluation (see Section 6). However, it is possible to adopt different strategies to combine the topics produced by the two modules, resulting in various trade-offs between precision and recall. Intuitively, the topics that get explicitly referred to in the text, returned by the syntactic module, tend to be more accurate, but including also the semantically related topics allows for a better recall. We will further discuss this in Section 6.1.

# 5 Creation of the Gold Standard

Since the CSO ontology was only released a few months ago, we lacked a dataset of manually annotated papers that could be used as gold standard. Therefore, we built such a gold standard by asking 21 domain experts to classify 70 papers in terms of topics drawn from the CSO ontology. This new gold standard has two objectives. First, it allows us to evaluate the proposed classifier against baseline methods, and, second, it provides a resource which will facilitate further evaluations in this area from other members of the research community.

## 5.1 Data preparation

We queried the MAG dataset and selected the 70 most cited papers published in 2007-2017 within the fields of *Semantic Web* (23 papers), *Natural Language Processing* (23 papers), and *Data Mining* (24 papers)[14].

We then contacted 21 researchers in these fields, at various levels of seniority, and asked each of them to annotate 10 of these papers. We structured the data collection in order to have each paper annotated by at least three experts, using majority vote to address disagreements. The papers were randomly assigned to experts, while minimising the number of shared papers between each pair of experts.

## 5.2 Data collection

We designed a web application to support the domain experts in annotating the papers. For each paper, the application displayed to the users: title, abstract, keywords (when available), and the set of candidate topics. The experts were asked to carefully read all the information and assess a set of candidate topics by dragging them in two different baskets: *relevant* and *not relevant*. They also could input further CSO topics that according to their judgment were missing from the candidate topics. Each paper was assigned with an average of $18 \pm 9$ topics.

We created the initial set of candidate topics by aggregating the output of three classifiers: the syntactic module (Section 4.1), the semantic module (Section 4.2), and a third approach, which was introduced for reducing the bias towards the first two methods. The latter first splits the input document into overlapping windows of size 10 (same as the training window of the word2vec model), each of them overlapping by 5 words. Then, for each window, it computes the average of the embedding vectors of all its words, creating an embedding representation of the window, and uses the word2vec model to identify the top 20 similar words with similarity above 0.6. It then assigns to each CSO concept a score based on the number of times it is found in the list of similar words and on the embedding similarity (cosine similarity between the vector representation of the window and word embedding). Finally, it sorts them in descending order and prunes the result set using the elbow method [27]. The combination of these approaches produced a very inclusive set of $41.8 \pm 17.5$ candidate topics for each paper.

---

[14] These three fields are well covered by CSO, which includes a total of 35 sub-topics for the Semantic Web, 173 for Natural Language Processing, and 396 for Data Mining.

### 5.3 Gold Standard

The data collection process produced 210 annotations (70 papers times 3 annotations per paper). In order to consider the taxonomic relationships of CSO, the resulting set of topics were semantically enriched by including also their direct super-areas as in [1, 2].

We computed the Fleiss' Kappa to measure the agreement among the three annotators on each paper. We obtained an average of 0.451 ± 0.177 indicating a moderate inter-rater agreement, according to Landis and Koch [28].

We created the gold standard using the majority rule approach. Specifically, if a topic was considered relevant by at least two annotators, it was added to the gold standard. Each paper in the gold standard is associated with 14.4 ± 7.0 topics.

## 6 Evaluation

We evaluated the CSO Classifier against thirteen alternative approaches on the task of classifying the papers in the gold standard according to CSO topics. Table 5 summarizes their main features and reports their performance.

**TF-IDF** returns for each paper a ranked list of words according to their TF-IDF score. The IDF of the terms was computed on the dataset of 4.6M papers in Computer Science, introduced in Section 4.2.1. **TF-IDF-M** maps these terms to CSO by returning all the CSO topics having Levenshtein similarity higher than 0.8 with them.

The following six classifiers use the Latent Dirichlet Allocation (LDA) [17] over the same corpus and then produce a number of keywords extracted from the distribution of terms associated to the LDA topics. **LDA100** was trained on 100 topics, **LDA500** on 500 topics, and **LDA1000** on 1000 topics. These three classifiers select all LDA topics with a probability of at least $j$ and return all their words with a probability of at least $k$. **LDA100-M**, **LDA500-M**, and **LDA1000-M** work in the same way, but the resulting keywords are then mapped to the CSO topics. In particular, they return all CSO topics that have Levenshtein similarity higher than 0.8 with the resulting set of terms. We performed a grid search for finding the best values of $j$ and $k$ on the gold standard and report here the best results of each classifier in term of f-measure.

**W2V-W** is the classifier described in Section 5.2 in order to produce further candidate topics for the domain experts. It processes the input document with a sliding window and uses the word2vec model to identify concepts semantically similar to the embedding of the window.

**STM** is the classifier originally adopted by Smart Topic Miner [2], the application used by Springer Nature for classifying proceeding in the field of Computer Science. It works similarly to the syntactic module described in Section 4.1, but it detects only exact matches between the terms extracted from the text and the CSO topics. **SYN** is the first version of the CSO classifier, originally introduced in [9], and it is equivalent to the syntactic module as described Section 4.1. **SEM** consists of the semantic module described in Section 4.2. **INT** is a hybrid version that returns the intersection of the topics produced by the syntactic (**SYN**) and semantic (**SEM**) modules. Finally, **CSO-C** is the default implementation of the CSO Classifier presented in this paper. As described in Section 4, it produces the union of the topics returned by the two modules.

We assessed the performance of these fourteen approaches by means of precision, recall and f-measure. When classifying a given paper $p$, the value of precision $pr(p)$ and recall $re(p)$ are computed as shown in Eq. 3:

$$\text{pr}(p) = \frac{|cl(p) \cap gs(p)|}{|cl(p)|} \qquad \text{re}(p) = \frac{|cl(p) \cap gs(p)|}{|gs(p)|} \qquad (3)$$

where *cl(p)* identifies the topics returned by the classifier, and *gs(p)* the gold standard obtained for that paper, including the super-areas of the gold standard used to enrich the user annotations as mentioned in Section 5.3. In order to obtain a better comparison between the different classifiers, we enhanced the results of each method with their direct super-concepts. The overall precision and recall for a given classifier are computed as the average of the values of precision and recall obtained over the papers. The f-measure (F1) is the harmonic mean of precision and recall.

Table 5. Values of precision, recall, and f-measure for the classifiers. In bold the best results.

| Classifier | Description | Prec. | Rec. | F1 |
|---|---|---|---|---|
| TF-IDF | TF-IDF. | 16.7% | 24.0% | 19.7% |
| TF-IDF-M | TF-IDF mapped to CSO concepts. | 40.4% | 24.1% | 30.1% |
| LDA100 | LDA with 100 topics. | 5.9% | 11.9% | 7.9% |
| LDA500 | LDA with 500 topics. | 4.2% | 12.5% | 6.3% |
| LDA1000 | LDA with 1000 topics. | 3.8% | 5.0% | 4.3% |
| LDA100-M | LDA with 100 topics mapped to CSO. | 9.4% | 19.3% | 12.6% |
| LDA500-M | LDA with 500 topics mapped to CSO. | 9.6% | 21.2% | 13.2% |
| LDA1000-M | LDA with 1000 topics mapped to CSO. | 12.0% | 11.5% | 11.7% |
| W2V-W | W2V on windows of words (*Section 5.2*). | 41.2% | 16.7% | 23.8% |
| STM | Classifier used by STM, introduced in [2]. | **80.8%** | 58.2% | 67.6% |
| SYN | Syntactic module (*Section 4.1*) [9]. | 78.3% | 63.8% | 70.3% |
| SEM | Semantic module (*Section 4.2*). | 70.8% | 72.2% | 71.5% |
| INT | Intersection of SYN and SEM. | 79.3% | 59.1% | 67.7% |
| CSO-C | The CSO Classifier. | 73.0% | **75.3%** | **74.1%** |

### 6.1 Results

We ran the fourteen classifiers and evaluated their results against the gold standard. In Table 5 we report the resulting values of precision, recall and f-measure.

The approaches based on LDA and TF-IDF performed poorly and did not exceed 30.1% of f-measure. It should be noted that while a tighter threshold on the Levenshtein similarity used for matching terms with CSO topics may further raise the precision, the low recall makes these approaches mostly unfit for this task. An analysis on the LDA topics showed that these tend to be mostly noisy and coarse-grained. They are thus unable to return several of the most specific CSO topics and often cluster together distinct CSO topics (e.g., "databases" and "search engines") in the same LDA topic. For instance, the LDA topic characterizing papers about Social Networks includes as top words many generic terms such as *users*, *online*, *social*, *profile*, *trust*, and so on. In general, LDA works quite well at identifying the main topics characterizing large collection of documents, but it is typically less suitable when trying to infer more specific research topics, which may be associated with a low number of publications (50-200), as discussed in [21]. **W2V-W** performed also poorly in term of both precision (41.2%) and recall (16.7%).

**STM** and **SYN** yielded a very good precision of respectively 80.8% and 78.3%. Indeed, these methods are good at finding topics that get explicitly mentioned in the text, which tend to be very relevant. However, they failed to detect some more subtle topics that are just implied, suffering from a low recall of 58.2% and 63.8%. The method used to map the terms from the text to the CSO topics plays a key role in the difference of performance between these two classifiers. Indeed, **STM** identifies only concepts that match exactly at least one of the terms extracted from the text. Conversely, **SYN** finds also partial matches, reducing precision but increasing recall.

The semantic module (**SEM**) lost some precision in comparison with **SYN**, but provided a better recall and f-measure. This suggests that it is able to identify further topics that do not directly appear directly in the paper, but naturally this may also produce some more false positives. **INT** yielded a higher precision (79.3%) compared to the syntactic and the semantic modules (78.3% and 70.8%), but it did not perform well in term of recall, which dropped from 63.8% and 72.2% to 59.1%.

Finally, the CSO Classifier (**CSO-C**) outperformed all the other methods in terms of both recall (75.3%) and f-measure (74.1%).

We compared the performance of the approaches using the McNemar's test for correlated proportions. The **CSO-C** performed significantly better ($p<10^{-7}$) than all the other approaches. In addition, STM [2], SYN [9], SEM, and INT were also significantly different from all the other baselines based on TF-IDF and LDA ($p<10^{-7}$). In summary, the CSO Classifier yielded the best overall results. However, it is possible to obtain a better precision by adopting a purely syntactic method that focus on the topics that are explicitly referred to in the text.

Another way to obtain a specific trade-off between precision and recall is changing the method used for selecting the returned topics from the ranked list produced by the semantic module. Intuitively, selecting the ones with the highest weights will yield a high precision, while being more inclusive will result in a higher recall. Figure 4 shows the value of precision, recall, and f-measure obtained by taking the first *n* topics in the ranked list. The precision (blue line) decreases while the recall (orange line) increases. The intersection of these two curves determines the highest value of f-measure (green line), with a peak of 63.6% when selecting the first 10 topics. It is useful to note that the elbow method, yielding a f-measure of 71.5%, clearly outperforms this solution based on a fixed number of returned topics.

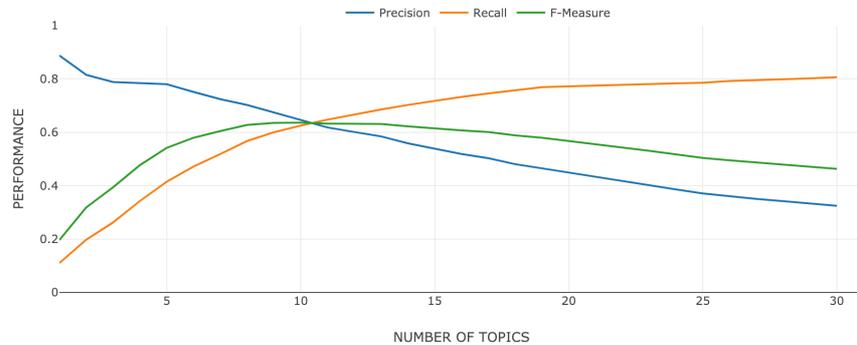

Figure 4. Average values of precision, recall, and f-measure according to the different sizes of candidate topic set returned for each paper.

As final step of the analysis, we treated the CSO Classifier as another expert and we observed how this influenced the inter-rater agreement. The general agreement when including CSO Classifier slightly lowers to 0.392 ± 0.144 yielding a moderate agreement with the majority of human experts.

# 7 Conclusions and future work

In this paper, we introduced the CSO Classifier, an application for classifying academic documents according to the Computer Science Ontology (CSO). The CSO Classifier analyses the text associated with research papers (title, abstract, and keywords) both on a syntactic and semantic level and returns a set of pertinent research topics drawn from CSO. This solution was evaluated on a gold standard of 70 manually annotated articles and outperformed the alternative approaches in terms of recall and f-measure. The code of the CSO Classifier and all the relevant material is freely available to the wider research community.

The approach presented in this paper opens up several interesting directions of work. On the research side, we will investigate further solutions combining natural language processing, machine learning, and semantics to improve the performance of the CSO Classifier. We also plan to explore the application of this approach to other research fields. In particular, we are currently working on a topic ontology for the Engineering field and we plan to produce a version of the classifier tailored to this area. We are also planning to extend it to the field of Medicine, in which we can take advantage of MeSH as ontology of subjects and the Medline dataset[15] for training our word2vec model.

On the technology transfer side, we will include the CSO Classifier within the pipelines of the Smart Topic Miner [2] and Smart Book Recommender [23], two applications we developed to support editorial processes at Springer Nature.

## References

1. Salatino, A.A., Osborne, F., Motta, E.: AUGUR: Forecasting the Emergence of New Research Topics. In: Joint Conference on Digital Libraries 2018, Fort Worth, Texas. pp. 1–10 (2018).
2. Osborne, F., Salatino, A., Birukou, A., Motta, E.: Automatic Classification of Springer Nature Proceedings with Smart Topic Miner. Semant. Web -- ISWC 2016. 9982 LNCS, 383–399 (2016).
3. Bolelli, L., Ertekin, Ş., Giles, C.L.: Topic and trend detection in text collections using latent dirichlet allocation. Lect. Notes Comput. Sci. (including Subser. Lect. Notes Artif. Intell. Lect. Notes Bioinformatics). 5478 LNCS, 776–780 (2009).
4. Griffiths, T.L., Steyvers, M.: Finding scientific topics. Proc. Natl. Acad. Sci. U. S. A. 5228–35 (2004).
5. Osborne, F., Motta, E.: Mining semantic relations between research aeas. In: The Semantic Web – ISWC 2012. pp. 410–426 (2012).
6. Salatino, A.A., Thanapalasingam, T., Mannocci, A., Osborne, F., Motta, E.: The Computer Science Ontology : A Large-Scale Taxonomy of Research Areas. In: The Semantic Web -- ISWC 2018. Springer (2018).
7. Song, Y., Roth, D.: Unsupervised Sparse Vector Densification for Short Text Similarity. In: Human Language Technologies: Annual Conference of the North American Chapter of the ACL. pp. 1275–80 (2015).
8. Lilleberg, J., Zhu, Y., Zhang, Y.: Support vector machines and Word2vec for text classification with semantic features. In: 2015 IEEE 14th International Conference on Cognitive Informatics & Cognitive Computing (ICCI*CC). pp. 136–140. IEEE (2015).
9. Salatino, A.A., Thanapalasingam, T., Mannocci, A., Osborne, F., Motta, E.: Classifying Research Papers with the Computer Science Ontology. In: ISWC-P&D-Industry-BlueSky 2018 (2018).
10. Decker, S.L., Aleman-meza, B., Cameron, D., Arpinar, I.B.: Detection of Bursty and

---

[15] Medline dataset: https://www.nlm.nih.gov/bsd/medline.html

Emerging Trends towards Identification of Researchers at the Early Stage of Trends. (2007).
11. Mai, F., Galke, L., Scherp, A.: Using Deep Learning for Title-Based Semantic Subject Indexing to Reach Competitive Performance to Full-Text. In: JCDL '18 Proceedings of the 18th ACM/IEEE on Joint Conference on Digital Libraries. pp. 169–178. ACM New York, Fort Worth, Texas, USA (2018).
12. Chernyak, E.: An Approach to the Problem of Annotation of Research Publications. In: Proceedings of the Eighth ACM International Conference on Web Search and Data Mining - WSDM '15. pp. 429–434. ACM Press, New York, New York, USA (2015).
13. Allan, J., Carbonell, J., Doddington, G., Yamron, J., Yang, Y.: Topic Detection and Tracking Pilot Study Final Report. (1998).
14. Osborne, F., Scavo, G., Motta, E.: Identifying diachronic topic-based research communities by clustering shared research trajectories. In: The Semantic Web: Trends and Challenges. pp. 114--129. Springer International Publishing (2014).
15. Small, H., Boyack, K.W., Klavans, R.: Identifying emerging topics in science and technology. Res. Policy. 43, 1450–67 (2014).
16. Caragea, C., Bulgarov, F., Mihalcea, R.: Co-Training for Topic Classification of Scholarly Data. Association for Computational Linguistics (2015).
17. Blei, D.M., Ng, A.Y., Jordan, M.I.: Latent Dirichlet Allocation. J. Mach. Learn. Res. 3, 993–1022 (2003).
18. Bhatia, S., Lau, J.H., Baldwin, T.: Automatic Labelling of Topics with Neural Embeddings. (2016).
19. Duvvuru, A., Radhakrishnan, S., More, D., Kamarthi, S.: Analyzing Structural & Temporal Characteristics of Keyword System in Academic Research Articles. Procedia - Procedia Comput. Sci. 20, 439–445 (2013).
20. Zhang, Y., Lu, J., Liu, F., Liu, Q., Porter, A., Chen, H., Zhang, G.: Does deep learning help topic extraction? A kernel k-means clustering method with word embedding. J. Informetr. 12, 1099–1117 (2018).
21. Osborne, F., Motta, E.: Klink-2: Integrating Multiple Web Sources to Generate Semantic Topic Networks. In: The Semantic Web - ISWC 2015. pp. 408–424 (2015).
22. Osborne, F., Motta, E., Mulholland, P.: Exploring scholarly data with rexplore. Semant. Web -- ISWC 2013. 8218 LNCS, 460–477 (2013).
23. Thanapalasingam, T., Osborne, F., Birukou, A., Motta, E.: Ontology-Based Recommendation of Editorial Products. In: International Semantic Web Conference 2018. , Monterey, CA (USA) (2018).
24. Narayanan, A., Shmatikov, V.: De-anonymizing Social Networks. In: 30th IEEE Symposium on Security and Privacy. pp. 173–187. IEEE (2009).
25. Mikolov, T., Chen, K., Corrado, G., Dean, J.: Efficient Estimation of Word Representations in Vector Space. (2013).
26. Mikolov, T., Chen, K., Corrado, G., Dean, J.: Distributed Representations of Words and Phrases and their Compositionality. In: Advances in neural information processing systems. pp. 3111–3119 (2013).
27. Satopää, V., Albrecht, J., Irwin, D., Raghavan, B.: Finding a "Kneedle" in a Haystack: Detecting Knee Points in System Behavior. In: ICDCSW '11 Proceedings of the 2011 31st International Conference on Distributed Computing Systems. pp. 166–171. IEEE Computer Society Washington (2011).
28. Landis, J.R., Koch, G.G.: The measurement of observer agreement for categorical data. Biometrics. 33, 159–74 (1977).